\documentclass[a4paper,11pt]{article}
\usepackage{pos}
\usepackage{here}

\title{Charged Higgs boson signatures via $pp \rightarrow H^\pm bj$ in 2HDM type-I}

\author[a]{R. Benbrik}
\author[b]{M. Krab}
\author[b]{B. Manaut}
\author*[a]{M. Ouchemhou}

\affiliation[a]{Polydisciplinary Faculty, Laboratory of Fundamental and Applied Physics, Cadi Ayyad University,\\ Sidi Bouzid, B.P. 4162, Safi, Morocco}

\affiliation[b]{Research Laboratory in Physics and Engineering Sciences, Modern and Applied Physics Team, Polydisciplinary Faculty, Beni Mellal, 23000, Morocco.}

\emailAdd{r.benbrik@uca.ac.ma}
\emailAdd{mohamed.krab@usms.ac.ma}
\emailAdd{b.manaut@usms.ma}
\emailAdd{mohamed.ouchemhou@ced.uca.ac.ma}

\abstract{In this study, we investigate the light charged Higgs boson production via $pp \rightarrow H^\pm bj$ in the type-I configuration of the Two-Higgs Doublet Model (2HDM). In the viable parameter space, assuming either $h$ or $H$ is the SM-like Higgs boson observed at the Large Hadron Collider (LHC), we focus on the bosonic decays $H^\pm \rightarrow W^\pm A/h$ and examine the final states arising from the above charged Higgs production and decay. We show here that the suggested signatures can have sizable rates, reaching the pb level, when the top quark is produced on-shell and $\tan\beta$ is small.}

\FullConference{%
  41st International Conference on High Energy physics - ICHEP2022\\
  6-13 July, 2022\\
  Bologna, Italy
}

\begin{document}
\maketitle
\section{Introduction}
The revolution of the standard model (SM) was made at the CERN LHC-collider in 2012 by discovering the Higgs boson predicted \cite{ATLAS:2012yve,CMS:2012qbp} with properties which are in great agreement with the SM prediction. However, there are many observations and theoretical predictions calling for new physics models outside the realm of the SM, such as the baryon asymmetry and matter-antimatter abundance. These new models come with a new symmetry, particles, and a new phenomenology. One of the simplest models is the Two-Higgs Doublet Model (2HDM), which includes, in addition to the SM-like Higgs boson $h$, a CP-even $H$, a CP-odd $A$, and two charged Higgs bosons $H^\pm$. The charged Higgs boson attracts the most interest since it is completely missing from the SM. Finding this particle at a collider would be a clear indication of a new era in physics. Direct searches for this charged Higgs boson(s) have already been conducted at LEP \cite{ALEPH:2013htx}, Tevatron \cite{CDF:2011pxh}, and now at the LHC \cite{ATLAS:2021upq,CMS:2020imj}, all yielding negative results. The injection of a soft broken $Z_2$ symmetry to forbid Flavor Changing Neutral Currents (FCNCs) at tree-level generates four versions of the model when it's applied to  the fermionic sector, denoted as: 2HDM type-I, type-II, type-X, and type-Y. Depending on these types, indirect constraints were set, namely the flavor observable dominated by the decay $B\to X_s\gamma$ allowing charged Higgs bosons with masses of less than 100 GeV in type-I and -X but excluding charged states with masses of less than 800 GeV in type-II and -Y. The aim of this study is to investigate the $H^\pm bj$ production \cite{Arhrib:2022ehv}, in the framework of 2HDM type-I, focusing on the signatures resulting from such a production and the subsequent bosonic decays, $H^\pm \rightarrow W^\pm A/h$.
This contribution is organized as follows. We briefly introduce  the 2HDM in section \ref{model}.
We discuss our results in section \ref{results} and we conclude in section \ref{conclusion}.
\section{The 2HDM}
\label{model}
2HDM is an extension of SM by another complex $SU(2)$ doublet. Its CP-conserving scalar potential with a softly broken $Z_2$ symmetry, which is $SU(2)_L \otimes U(1)_Y$ invariant, is given as follows
\begin{align}
V_{\rm{2HDM}} ~&=~ m_{11}^2(\Phi_1^+\Phi_1)+m_{22}^2(\Phi_2^+\Phi_2)-m_{12}^2(\Phi_1^+\Phi_2+ \rm{h.c.})+\lambda_1(\Phi_1^+\Phi_1)^2+\lambda_2(\Phi_2^+\Phi_2)^2\nonumber \\
&~ +\lambda_3(\Phi_1^+\Phi_1)(\Phi_2^+\Phi_2)
+\lambda_4 (\Phi_1^+\Phi_2)(\Phi_2^+\Phi_1)+\frac{\lambda_5}{2}[(\Phi_1^+\Phi_2)^2+ \rm{h.c.}], \label{RTHDMpot}
\end{align}
where the parameters $m_{11}^2, m_{22}^2, m_{12}^2$ and $\lambda_{1-5}$ are real. Each doublets $\Phi_{1,2}$ developed a vacuum expectation values $v_{1,2}$, therefore the model becomes under the control of the following parameters: $m_{11}^2, m_{22}^2, m_{12}^2, v_{1,2}$ and $\lambda_{1-5}$. However, we could fall within seven free and real-independent parameters by introducing the minimization conditions of the potential. These parameters, $\{m_h,~m_H,~m_A,~m_{H^\pm},~\alpha,~\tan\beta,~m_{12}^2 \}$, are chosen on a physical basis for a realistic description of the model. Here, we are interested in 2HDM type-I, in which $\Phi_2$ couples to all fermions like in the SM.
\section{Results and discussion}
\label{results}
\begin{figure}[H]
\centering
\includegraphics[scale=0.3]{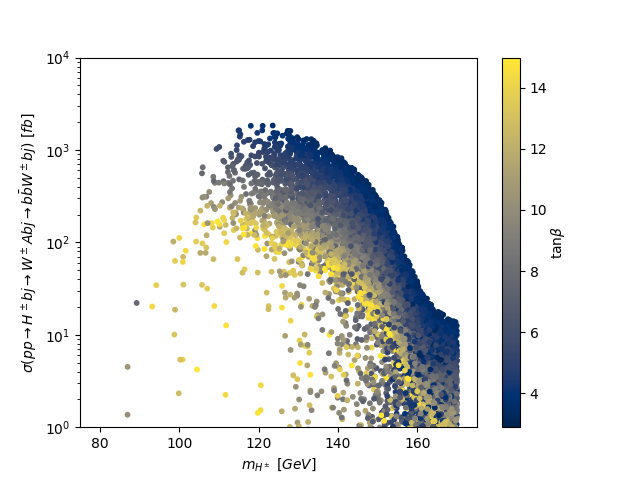} 
\includegraphics[scale=0.3]{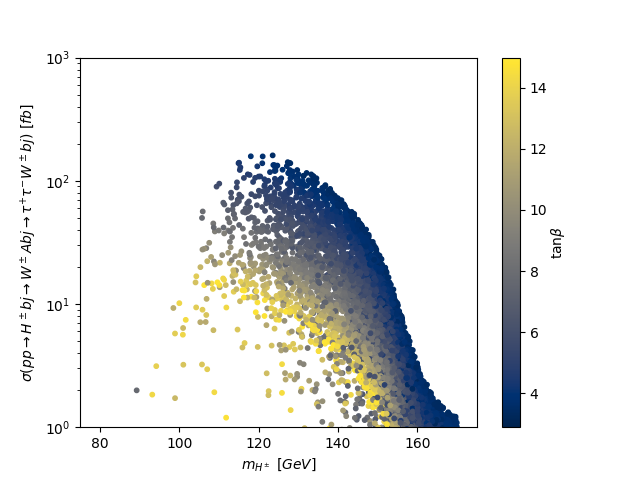} 
\includegraphics[scale=0.3]{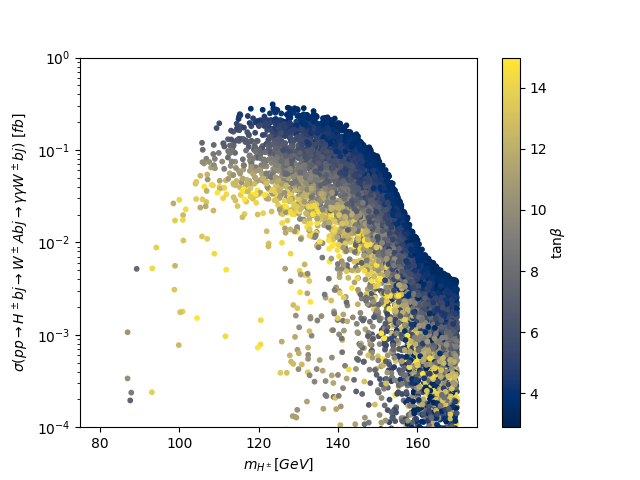}\\
\includegraphics[scale=0.3]{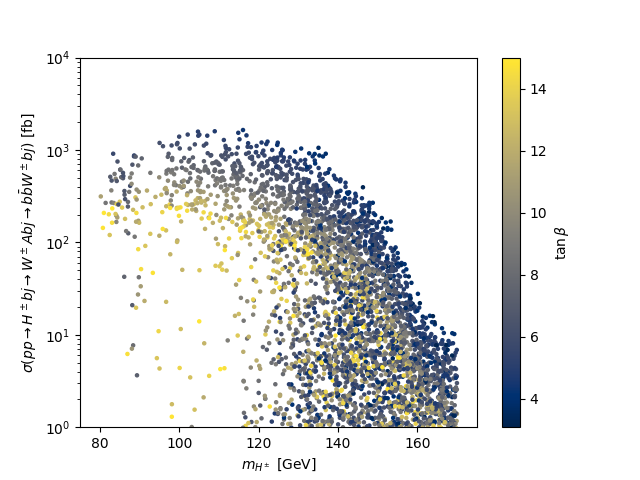}	
\includegraphics[scale=0.3]{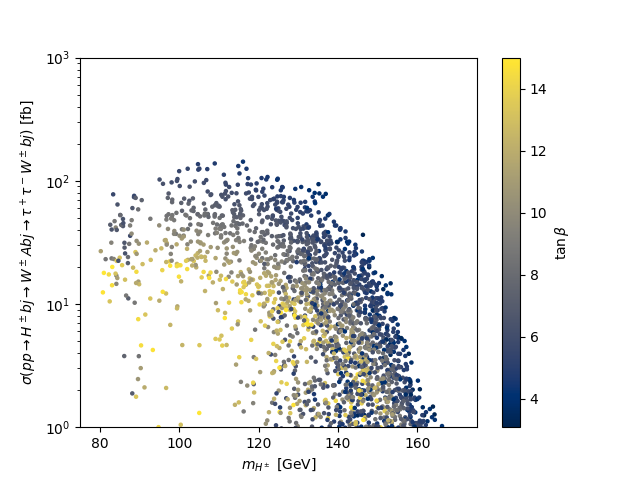}
\includegraphics[scale=0.3]{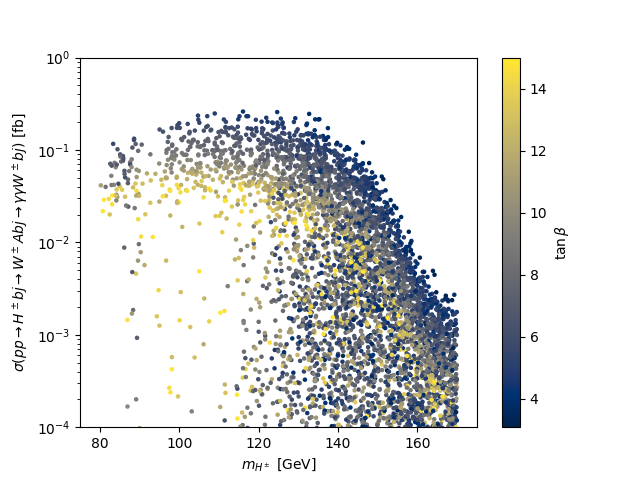}
\caption{$\sigma(pp \rightarrow b\bar{b}W^\pm bj)$ (left panels), $\sigma(pp \rightarrow \tau^+ \tau^- W^\pm bj)$ (middle panels) and $\sigma(pp \rightarrow \gamma\gamma W^\pm bj)$ (right panels) as a function of $m_{H^\pm}$, with the color code indicating $\tan\beta$. Upper (lower) panels present the SH (IH) results.} \label{xs_HcW}
\end{figure}

The 2HDM parameters are randomly scanned, in both Standard Hierarchy (SH) and Inverted Hierarchy (IH), as illustrated in Table \ref{param_scans}.
During the scan, we take into account the theoretical constraints, such as vacuum stability, perturbativity, and unitarity pertubative as well as the available experimental constraints, including collider bounds, Higgs signal strength, EWPOs through the oblique parameters ($S$, $T$, and $U$), and B-physics observables. Note that the scan is performed using the combination of the following public tools: 2HDMC, HiggsBounds, and HiggsSignals as well as SuperIso, to check the theoretical and experimental constraints mentioned above. 
\begin{table}[H]
	\centering
	\renewcommand{\arraystretch}{1.4} %
	\setlength{\tabcolsep}{1.5pt}
	\begin{tabular}{|c|c|c|c|c|c|c|c|}\hline
		& $m_h$ & $m_H$ & $m_A$ &	$m_{H^\pm}$ & $s_{\beta-\alpha}$ & $\tan\beta$ & $m_{12}^2 $ \\\hline
		SH &  $125.09$ & $[126,\,200]$ & $[60,\,200]$ & $[80,\,170]$ & $[0.95,\,1]$ & $[2,\,15]$ & $[0,\,m_H^2\cos\beta\sin\beta]$\\\hline
		IH & $[10,\,120]$ & $125.09$ & $[60,\,200]$ & $[80,\,170]$ & $[-1,\,1]$ & $[2,\,15]$ & $[0,\,m_h^2\cos\beta\sin\beta]$\\\hline
	\end{tabular}		
	\caption{2HDM input parameters adopted. All masses are in GeV.} \label{param_scans}
\end{table}
Considering $H^\pm \to W^\pm A$, the upper panels of Fig. \ref{xs_HcW} show the signatures $b\bar{b}W^\pm bj$ (left panel), $\tau^+ \tau^- W^\pm bj$ (medium panel) and $\gamma\gamma W^\pm bj$ (right panel) as a function of $m_{H^\pm}$, in the SH. As illustrated in this figure, these final states are potentially important for smaller $\tan\beta$, reaching near $2$ pb for $b\bar{b}W^\pm bj$ and $170$ fb for $\tau^+ \tau^- W^\pm bj$.  Similar to upper panels, we show in the lower panels the same final states but in the IH. Our signatures in this case are of the same order of magnitude as in the SH. We found the rate for $\gamma\gamma W^\pm bj$ is negligible in both standard and inverted hierarchies.

\begin{figure}[H]	
\includegraphics[scale=0.3]{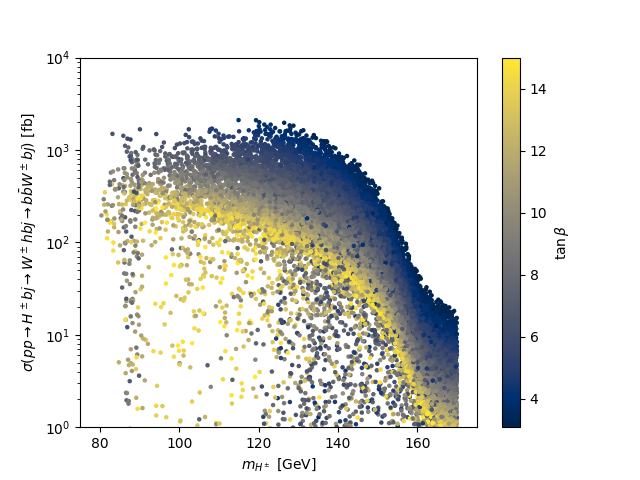}
\includegraphics[scale=0.3]{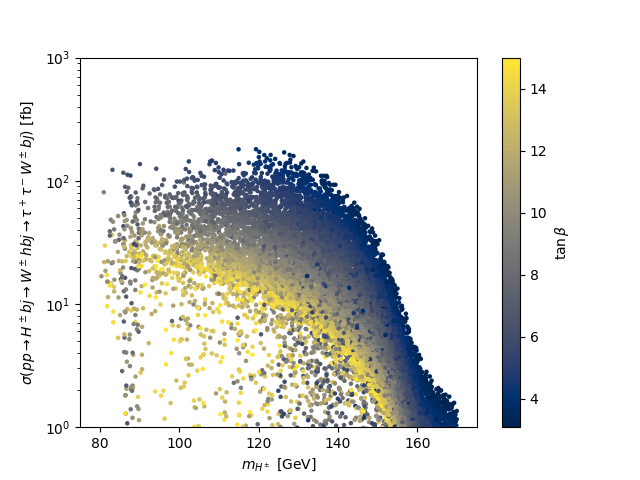}
\includegraphics[scale=0.3]{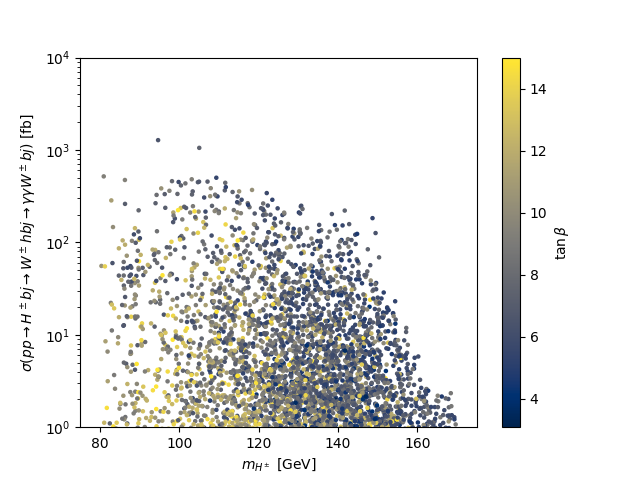}
\caption{$\sigma(pp \rightarrow b\bar{b}W^\pm bj)$ (left panel), $\sigma(pp \rightarrow \tau^+ \tau^- W^\pm bj)$ (middle panel) and $\sigma(pp \rightarrow \gamma\gamma W^\pm bj)$ (right panels) as a function of $m_{H^\pm}$, with the color code indicating $\tan\beta$.} \label{signatures3}
\end{figure}

In contrast to SH, the decay $H^\pm \to W^\pm h$ may also be dominant in IH and gives rise to the same signatures present above in Fig. \ref{signatures3}. It is clear that $b\bar{b}W^\pm bj$ (left panel) and $\tau^{+}\tau^{-}W^\pm bj$ (medium panel) are quite similar to the former case (Fig. \ref{xs_HcW}), with exception of the $\gamma\gamma W^\pm bj$ signature (right panel) that is potentially important and could reach the pb level in some cases.

\section{Conclusion}
\label{conclusion}
The sign for a new physics is the subject of recent and future collider experiments. In this study, we briefly investigate the charged Higgs boson production in association with a bottom quark and a jet in 2HDM type-I. Focusing on the $H^\pm \rightarrow W^\pm A/h$ decays, we have found an interesting rate for the $b\bar{b}W^\pm bj$, $\tau^+ \tau^- W^\pm bj$ and $\gamma\gamma W^\pm bj$ signatures that could enhance the LHC search for a light charged Higgs boson.

\bibliographystyle{JHEP}
\bibliography{my-bib-ref}
\end{document}